%% file: main.tex
\UseRawInputEncoding
\documentclass[10pt,journal]{IEEEtran}
\input{sections/Preamble}
\graphicspath{{./figs/}}
\begin{document}
\input{sections/title}
\input{sections/introduction}
\input{sections/preliminary}
\input{sections/motivation}

\input{sections/simulations}
\input{sections/Conclusions}
\bibliographystyle{references/IEEEtran}
\bibliography{main.bbl}
\end{document}

%% file: sections/Preamble.tex
\IEEEoverridecommandlockouts
\usepackage{cite}
\usepackage{amsmath,amsfonts,amsthm,bm,amssymb} %
\usepackage[table,xcdraw]{xcolor}
\usepackage{algorithm}
\usepackage{algpseudocode} 
\usepackage{graphicx}
\usepackage{graphics}
\usepackage{epsfig}
\usepackage{footnote}
\usepackage{textcomp}
\PassOptionsToPackage{no-math}{fontspec}
\usepackage{xcolor}
\usepackage{tikz}  
\usepackage{makecell}
\usepackage{bm}
\usetikzlibrary{calc} 
\usetikzlibrary{arrows,shapes,chains} 
\usetikzlibrary{positioning, shapes.geometric}
\usepackage{multirow} 
\usepackage{verbatim}
\usepackage{enumitem}
\def\BibTeX{{\rm B\kern-.05em{\sc i\kern-.025em b}\kern-.08em
		T\kern-.1667em\lower.7ex\hbox{E}\kern-.125emX}}
	
\usepackage{verbatim}
\usepackage{pgfplots}
\usepackage{hyperref}
\pgfplotsset{width=10cm,compat=1.9}
\pgfplotsset{
	every axis legend/.style={
		cells={anchor=center},
		inner xsep=3pt,inner ysep=2pt,
		nodes={inner sep=2pt,text depth=0.1em},
		anchor=north east,
		shape=rectangle,
		fill=white,draw=green,
		font=\footnotesize
	},
}

\usepackage{verbatim}
\usepackage{booktabs}
\usepackage{subcaption}
\usepackage{fancyhdr}
\usepackage{svg}
\normalsize
	

%% file: sections/title.tex
\title{Quasi-Belief Propagation and Neural-Network Check Node Processing for BCH Codes \\
}

\author{Guangwen Li
\thanks{G.Li is with School of Information \& Electronics, Shandong Technology and Business University, Yantai, China e-mail: lgw.frank@sdtbu.edu.cn}
}
\maketitle
\begin{abstract}
This paper proposes a quasi-BP decoding scheme for BCH codes that preserves the parallelizable structure of belief propagation while exploiting code automorphisms and optimized redundant parity-check matrices. To eliminate the computationally expensive $\tanh$ and $\tanh^{-1}$ functions in check node updates, we further introduce a neural-network-based variant that replaces them with a lightweight  convolutional neural network trained under a triple-constraint loss function enforcing non-negativity and order consistency. Simulation results for three BCH codes demonstrate that quasi-BP decoding achieves competitive frame error rate performance, with a gap within 0.25 decibels compared with belief propagation decoding of an LDPC code of similar blocklength. The neural-network-based variant incurs negligible performance loss while enabling stable deployment with arithmetic operations on hardware accelerators. Concatenation with an ordered statistics decoding variant further bridges the gap to the maximum-likelihood bound. Hence, the proposed schemes offer a viable path toward high-throughput, low-latency decoding of BCH codes in next-generation communication systems.
\end{abstract}



%% file: sections/introduction.tex
\section{Introduction}
\IEEEPARstart{S}{ince} Shannon's seminal work \cite{shannon1948mathematical}, channel coding has been a cornerstone of reliable communication. Among linear block codes, low-density parity-check (LDPC) codes stand out for their exceptional error-correcting performance and their ability to approach the Shannon limit under belief propagation (BP) decoding \cite{gallager62, mackay96}. In complexity-constrained implementations, BP is often replaced by min-sum (MS) variants \cite{zhao05}, such as the normalized MS (NMS) algorithm \cite{chen2005reduced}. However, these BP-based methods are suboptimal for finite-length LDPC codes due to short cycles in their Tanner graphs, and the error floor phenomenon \cite{richardson2003error} in the high signal-to-noise ratio (SNR) regime further complicates the design of the parity-check matrix (PCM) $\mathbf{H}$.

In contrast, classical high-density parity-check (HDPC) codes, including Bose--Chaudhuri--Hocquenghem (BCH)  codes, are widely adopted in storage and deep-space communications, owing to their large minimum distances and efficient hard-decision decoding algorithms. Yet, soft BP decoding performs poorly on BCH codes, as their algebraic structure inevitably introduces numerous short cycles into the Tanner graph. Hence, a key challenge in repurposing BCH codes for next-generation systems---such as the Internet of Things (IoT) \cite{zhang2024improved}---is to design a decoding scheme that matches LDPC BP decoding in terms of frame error rate (FER), computational complexity, throughput, and latency.

Ordered statistics decoding (OSD) \cite{Fossorier1995} was introduced to bridge the performance gap between BP and maximum-likelihood (ML) decoding for short codes. Recent efforts to accelerate OSD for BCH and short LDPC codes \cite{Yue2021} have focused on early stopping criteria or reducing the number of test error patterns via thresholding. Despite these improvements, the inherently sequential nature of OSD and its variants limits their suitability for high-throughput or low-latency applications.

The rise of deep learning has opened new avenues for error-correction coding. Nachmani et al. \cite{nachmani16} pioneered the unfolding of BP iterations into a neural network (NN), yielding neural belief propagation (NBP) with trainable edge weights. Other NN architectures, including convolutional neural networks (CNNs) and recurrent neural networks (RNNs), have also been investigated for various linear codes \cite{nachmani18, liang18}. A two-stage decimation approach \cite{buchberger21} first identifies likely bits via an NN and then performs list decoding with NBP, albeit at the cost of sacrificing the parallel processing capability of BP. In a different direction, \cite{von2024spiking} proposed using spiking neural networks to approximate the check node updates of BP decoders for certain LDPC codes, targeting energy-efficient signal processing.

In our previous work \cite{li2025effective}, we showed that enhanced NMS (ENMS) can effectively decode BCH codes by exploiting the redundancy in $\mathbf{H}$ and the cyclic structure of the code. The present paper extends this idea to a canonical BP-like framework to investigate the fundamental limits of decoding when the channel noise variance is known. Our main contributions are as follows:

\begin{itemize}
    \item We propose a parallelizable quasi-BP (QBP) decoder for BCH codes that achieves competitive FER performance relative to standard BP decoding of LDPC codes in the medium-to-high SNR region.
    \item We replace the set of complex equations at each check node with a CNN-based model involving only arithmetic operations, yielding the QBP-SF decoder with negligible performance loss.
    \item Extensive simulations validate the proposed QBP and QBP-SF decoders for high-rate short BCH codes, and their concatenation with an OSD variant offers a viable path toward approaching the ML bound.
\end{itemize}

The remainder of this paper is organized as follows. Section~\ref{preliminary} reviews preliminaries on BP and OSD variants. Section~\ref{sec:motivations} details the motivations and formulation of the QBP and QBP-SF decoding methods. Section~\ref{simulations} presents FER results for various BCH codes, along with a brief complexity analysis and discussion. Finally, Section~\ref{conclusions} concludes the paper.

%% file: sections/preliminary.tex
\section{Preliminaries}
\label{preliminary}

Consider a binary message row vector $\mathbf{m} = [m_i]_{i=1}^K$ encoded into a codeword $\mathbf{c} = [c_i]_{i=1}^N$ via $\mathbf{c} = \mathbf{m}\mathbf{G}$ over GF(2), where $K$ and $N$ denote the message and codeword lengths, respectively, and $\mathbf{G}$ is a full-rank generator matrix.

Each codeword bit $c_i$ is mapped to an antipodal symbol using binary phase shift keying (BPSK) modulation: $s_i = 1 - 2c_i$. After transmission over an additive white Gaussian noise (AWGN) channel with noise $n_i \sim \mathcal{N}(0,\sigma^2)$, the received soft sequence $\mathbf{y} = [y_i]_{i=1}^N$, where $y_i = s_i + n_i$, is fed into a dedicated decoder to estimate the transmitted codeword. Since the log-likelihood ratio (LLR) for the $i$-th bit is derived as
\begin{equation}
	l_{i,{\text{ch}}} = \log\frac{p(y_i\mid c_i=0)}{p(y_i\mid c_i=1)} = \frac{2y_i}{\sigma^2},
	\label{eq:llr}
\end{equation}
a larger magnitude of $y_i$ implies higher reliability of the associated codeword bit.

\subsection{BP Variants and Neural Counterparts}

A code is defined by a bipartite Tanner graph underlying its PCM $\mathbf{H}$. This graph contains $N$ variable nodes and $M \geq N-K$ check nodes if redundancy is allowed, with an edge connecting variable node $j$ to check node $i$ whenever $H_{ij} \neq 0$.

Assuming a flooding schedule with at most $I_m$ iterations, standard BP exchanges LLR messages along the Tanner graph edges. The variable-to-check message at iteration $t$ ($t = 1,\dots,I_m$) is
\begin{equation}
	l_{\nu_i \to c_j}^{(t)} = l_{i,{\text{ch}}} + \sum_{\substack{ q \in \mathcal{C}(i)\setminus j}} l_{c_q \to \nu_i}^{(t-1)},
	\label{eq:v2c}
\end{equation}
where $\mathcal{C}(i)\setminus j$ denotes the set of neighboring check nodes of variable node $\nu_i$ except $c_j$, and $l_{c_q \to \nu_i}^{(0)} = 0$ is initialized. The check-to-variable message flows in the opposite direction as
\begin{equation}
	l_{c_j \to \nu_i}^{(t)} = 2\tanh^{-1}\!\Biggl(
	\prod_{\substack{ q \in \mathcal{V}(j)\setminus i}}
	\tanh\!\Bigl(\frac{l_{\nu_q \to c_j}^{(t)}}{2}\Bigr)
	\Biggr),
	\label{eq:c2v}
\end{equation}
where $\mathcal{V}(j)\setminus i$ denotes the set of neighboring variable nodes of check node $c_j$ except $\nu_i$.

After each iteration, the a posteriori LLR of the $i$-th bit is
\begin{equation}
	l_{\nu_i}^{(t)} = l_{i,{\text{ch}}} + \sum_{\substack{ q \in \mathcal{C}(i)}} l_{c_q \to \nu_i}^{(t-1)},
	\label{eq:aposteriori}
\end{equation}
which yields the tentative hard decision
\begin{equation}
	\widehat{c}_i^{(t)} = 
	\begin{cases}
		0, & \operatorname{sgn}\!\bigl(l_{\nu_i}^{(t)}\bigr)=+1,\\[2pt]
		1, & \text{otherwise}.
	\end{cases}
	\label{eq:hard_decision}
\end{equation}
Decoding terminates early to reduce complexity if $\widehat{\mathbf{c}}^{(t)}\mathbf{H}^{\mathsf T}=\mathbf{0}$, where $\widehat{\mathbf{c}}^{(t)}=[\widehat{c}_i^{(t)}]_{i=1}^N$. Otherwise, $t \leftarrow t + 1$ and another iteration is executed through \eqref{eq:v2c}--\eqref{eq:hard_decision} until the $I_m$-th iteration is reached.

To avoid the computationally expensive $\tanh$ and $\tanh^{-1}$ operations in \eqref{eq:c2v}, the MS approximation replaces \eqref{eq:c2v} with
\begin{equation}
	l_{c_j \to \nu_i}^{(t)} = s_{c_j \to \nu_i}^{(t)}\; \phi_{c_j \to \nu_i}^{(t)},
	\label{eq:ms}
\end{equation}
where
\begin{align}
	s_{c_j \to \nu_i}^{(t)} &= \prod_{\substack{ q \in \mathcal{V}(j)\setminus i}}
	\operatorname{sgn}\!\bigl(l_{\nu_q \to c_j}^{(t)}\bigr),\label{eq:sign}\\
	\phi_{c_j \to \nu_i}^{(t)} &= \min_{\substack{ q \in \mathcal{V}(j)\setminus i}}
	\bigl|l_{\nu_q \to c_j}^{(t)}\bigr|. \label{eq:min}
\end{align}
To mitigate the consistent overestimation inherent in the MS approximation, a multiplicative scaling factor is applied to \eqref{eq:min}, yielding the NMS decoder.

When trainable weights are assigned to or shared among the edges of a specific network trellis for any BP variant, the resulting neural decoder is referred to as NBP \cite{nachmani18}.

\subsection{OSD and Its Variants}

OSD variants \cite{Fossorier1995} refer to a family of universal decoding methods that approximate the ML bound by generating and selecting candidate test error patterns (TEPs). The procedure of traditional order-$p$ OSD consists of four steps. First, the received sequence is sorted by symbol reliability, and the corresponding permutation is applied to the PCM, which is then transformed into systematic form via Gaussian elimination; the resulting permutation is then applied to the sorted sequence. The reordered codeword is divided into the most reliable basis (MRB) and the least reliable basis, and all TEPs with Hamming weight up to a predetermined $p$ are generated on the MRB. Third, a candidate codeword is reconstructed by XORing the hard decision of the MRB with each TEP and then using the systematic PCM to recover the full codeword. Fourth, the candidate codeword that minimizes the weighted Hamming distance is selected as the final output.

Compared with traditional OSD, the enhanced OSD variant \cite{li2025neural} constructs an evolutionary decoding path based on the ALMLT algorithm \cite{kabat2007new} prior to decoding. It then boosts the reliability measurement of codeword bits via a small NN model before invoking OSD to inspect the TEP batch along the decoding path. Meanwhile, an early termination mechanism is facilitated by another NN model. As a result, the computational complexity and latency are substantially reduced without sacrificing error correction performance.

%% file: sections/motivation.tex
\input{sections/motivation_one}
\input{sections/motivation_two}

%% file: sections/motivation_one.tex
\section{Motivations and Methods}
\label{sec:motivations}

As a class of HDPC codes, BCH codes possess dense PCMs, which inherently introduce numerous short cycles in their Tanner graphs. These short cycles are widely recognized as the primary cause of performance degradation for BP and NMS decoders, owing to the correlated nature of iterative messages they induce. To mitigate this detrimental effect, the ENMS exploits both the redundancy of an optimized PCM and the automorphisms arising from the cyclic property of the code.

It is well known that NMS decoding can approach standard BP performance for most LDPC codes. However, its success is partly attributed to the limited number of variable nodes per check node involved in the check node update equations \eqref{eq:c2v} or \eqref{eq:ms}. In contrast, the check node update of ENMS involves a significantly larger number of variable nodes, but the minimization operation drastically reduces complexity by oversimplifying the strong check constraint. This effectively discards the diversified values that would otherwise be returned by \eqref{eq:c2v}, potentially degrading decoding performance. Motivated by these observations, we propose a QBP decoder that preserves the strengths of both ENMS and standard BP, as detailed below.

\subsection{QBP Iterative Decoding}

The decoding procedure of QBP comprises the following steps:

\begin{itemize}
	\item \textbf{Preprocessing:} Specify a factor $\delta_1$ to control the extent of input dilation; dilate and optimize the traditional $\mathbf{H}$ into $\mathbf{H}_o$ with a redundancy factor $\delta_2$. Interested readers are referred to \cite{li2025neural} for details on assigning the grouping of $(\delta_1,\delta_2,I_m)$.
	
	\item \textbf{Initialization:} Set $l_{\nu_i}^{(0)} = l_{i,\text{ch}}$ from \eqref{eq:llr}; initialize all check-to-variable messages $m_{c_j \rightarrow \nu_i}^{(0)} = 0$; set iteration index $t = 1$.
	
	\item \textbf{Variable node update:} This consists of three phases.
	
	\textit{Phase I -- Parsing and merging:} Given incoming messages $m_{c_j \rightarrow \nu_i}^{(t-1)}$, split and align them according to the indices of the original codeword bits, yielding intermediate message blocks $\tilde{m}_{c_j \rightarrow \nu_i}^{(t-1,w)}$, $w = 1, 2, \ldots, \delta_1$. The pre-dilation input is refreshed by combining the previous input with the result of merging $\tilde{m}_{c_j \rightarrow \nu_i}^{(t-1,w)}$ according to
	\begin{equation}
		l_{\nu_i}^{(t)} = l_{\nu_i}^{(t-1)} + \eta \sum_{w = 1}^{\delta_1} \sum_{j \in \mathcal{C}(i)} \tilde{m}^{(t-1,w)}_{c_j \rightarrow \nu_i},
		\label{quasi_bp_variable_node_update}
	\end{equation}
	where the aggregation coefficient $\eta$ can be optimized by minimizing the FER via a dichotomous line search.
    
	\textit{Phase II -- Normalization:} The message vector $\boldsymbol{l}_\nu^{(t)} = [l_{\nu_i}^{(t)}]_{i=1}^N$ is normalized as follows:
	\begin{equation}
	\boldsymbol{l}_\nu^{(t)} \longleftarrow \frac{2}{\sigma^2 m_v} \boldsymbol{l}_\nu^{(t)},
	\end{equation}
	where $m_v$ is the mean of the magnitudes of all components of $\boldsymbol{l}_\nu^{(t)}$.		

	\textit{Phase III -- Input dilation:} Leverage the automorphisms \cite{li2025neural} of the BCH codes to dilate the input $l_{\nu_i}^{(t)}$ by a factor $\delta_1$ into a block $L_{\nu_i}^{(t)}$, and dispatch the resulting values to the associated check nodes. That is, for each variable node $\nu_i$ in the dilated block,
	$
		m_{\nu_i \rightarrow c_j}^{(t)} = L_{\nu_i}^{(t)}, \quad j \in \mathcal{C}(i).
	$
   
	\item \textbf{Check node update:} Compute $m_{c_j \rightarrow \nu_i}^{(t)}$ identically to the standard BP update in \eqref{eq:c2v}.
	
	\item \textbf{Termination and hard decision:} Execute the remaining operations identically to the aforementioned BP or NMS after forming a tentative hard decision on $l_{\nu_i}^{(t)}$ from \eqref{quasi_bp_variable_node_update}.
\end{itemize}

The normalization imposed on the updated input offers several merits. First, it forces the expectation of $l_{\nu_i}^{(t)}$ to approach (or inherit) that of $l_{\nu_i}^{(0)}$, aligning with the key fact that the latter is invoked only once in QBP, which is a fundamental departure from conventional BP decoding. Second, the resulting message magnitudes are more stable, which brings twofold benefits: the FER performance becomes less sensitive to variations in $\eta$, and, more importantly, it enables feasible fitting of a group of check node update functions such as \eqref{eq:c2v} by an NN model as detailed below.

%% file: sections/motivation_two.tex
\subsection{NN Implementation of Check Node Equations}

For any check node $j$, let $|\mathcal{V}(j)| = r_w$ denote the number of non-zero elements per row of $\mathbf{H}_o$. Examining \eqref{eq:c2v} more closely, when the magnitudes of $l_{\nu_q \to c_j}^{(t)}$, $q \in \mathcal{V}(j)$, are sorted in descending order, the corresponding output group $l_{c_j \to \nu_q}^{(t)}$ will be in ascending order of magnitude. This follows directly from the fact that both $\tanh$ and $\tanh^{-1}$ are monotonically increasing functions. Therefore, to circumvent the complexity of \eqref{eq:c2v}, fitting an NN substitution for the group of update equations associated with the same check node is preconditioned on two actions that effectively shrink the search space: (i) separation of sign and magnitude for all $l_{\nu_q \to c_j}^{(t)}$, $q \in \mathcal{V}(j)$, as done by BP and its variants; and (ii) sorting of the magnitude parts.

\subsubsection{NN Architecture}

\begin{figure}[htbp]
	\centering	\includegraphics[width=\linewidth]{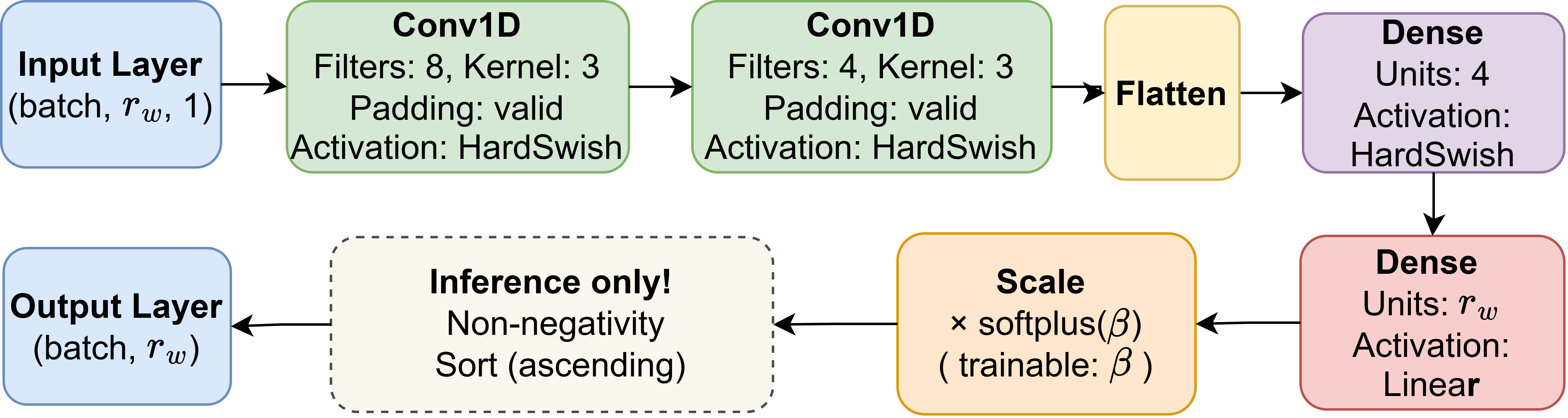}
	\caption{Architecture of NN substitution for a group of check node update equations of QBP decoding for the BCH(127,64) code, where the dotted block is skipped during training and applied only in the inference phase.}
	\label{fig:nn_substitution}
\end{figure}

The proposed NN, as illustrated in Fig.~\ref{fig:nn_substitution} for the BCH(127,64) code \cite{channelcodes}, is a lightweight 1D CNN with five trainable layers: two Conv1D layers for feature extraction, followed by a flatten operation, a Dense bottleneck layer, and a final Dense output layer of the same dimension $r_w$ as the input. A trainable scaling factor $\beta$ is applied to accelerate training convergence and will be merged with the coefficient $\eta$ in \ref{quasi_bp_variable_node_update} during inference. The non-linear activation function HardSwish is adopted due to its superior validation performance.

For longer codes with larger $r_w$, the number of layers or filters per layer must be increased to meet the fitting challenge. Alternatively, other solutions may include fitting either $\prod_{q \in \mathcal{V}(j)} \tanh(\cdot)|_q$ per check node $j$, or $\prod_{q \in \mathcal{V}(j)\setminus i} \tanh(\cdot)|_q$ directly between check node $j$ and variable node $i$; both approaches can substantially reduce the NN size when the output reduces to a scalar.

\subsubsection{Training Logistics}

Assume $\mathbf{H}_o$ has regular check degree $r_w$. For each check node, the total $r_w$ variable-to-check messages involved in \eqref{eq:c2v} of QBP decoding yield $r_w$ corresponding distinct check-to-variable messages, which are recorded as raw input and output vectors, respectively. Descending sorting of the input vector magnitudes and ascending sorting of the output vector magnitudes form one labeled sample. Suppose the maximum number of QBP iterations is $I_m$, $\mathbf{H}_o$ has $M$ rows, $S$ test points are uniformly selected within the SNR range of interest, and $X$ received sequences are applied at each test point; then a total of $I_m M S X$ samples can be obtained.

A triple-constraint loss function $\mathcal{L}_{\text{total}}$ simultaneously supervises three dimensions of the output values during training:
\begin{equation}
\mathcal{L}_{\text{total}} = \mathcal{L}_{\text{d}} + \mathcal{L}_{\text{o}} + \mathcal{L}_{\text{p}},
\end{equation}
with:
\begin{itemize}
	\item $\mathcal{L}_{\text{d}}$, supervising the closeness between predicted values and true labels, is defined as $\mathcal{L}_{\text{d}} = \sum_{b_s} \sum_{i=1}^{r_w} ( o_i - u_i )^2$, where $b_s$ denotes the batch size, and $o_i$ and $u_i$ are the $i$-th dimensions of the model output and label, respectively.

	\item $\mathcal{L}_{\text{o}}$, enforcing ascending order on the output vector, is defined as $\mathcal{L}_{\text{o}} = \sum_{b_s} \sum_{i=1}^{r_w-1} \text{ReLU}(o_i - o_{i+1})$.

	\item $\mathcal{L}_{\text{p}}$, enforcing all output values to be non-negative, is defined as $\mathcal{L}_{\text{p}} = \sum_{b_s} \sum_{i=1}^{r_w} \text{ReLU}(-o_i)$.
\end{itemize}

Taking the BCH (127,64) code as an example, the generated training sample size is approximately $10^6$. During training, the batch size $b_s$ is set to $200$, and the Adam optimizer is used to minimize the loss function $\mathcal{L}_{\text{total}}$. The initial learning rate $l_{ri}$ is set to $10^{-3}$, decayed by a factor of $0.995$ every $200$ steps, and training terminates after $l_{rt}=10^5$ steps.

\subsection{Fine-Tuning of the QBP-SF}

For the training samples provided at each specific SNR point, the QBP-SF, termed after replacing check node update operations with the trained NN, can be further fine-tuned by including the trainable $\eta$ as well, aiming to obtain optimal parameter values customized to the current SNR. The training settings are inherited from the NN training described above, except that $(b_s, l_{ri}, l_{rt}) = (30, 10^{-5}, 10^3)$ when calling the Adam optimizer to minimize the cross-entropy loss function.

%% file: sections/simulations.tex
\section{Simulation and Analysis}
\label{simulations}

The codes BCH(127,64), BCH(127,99), BCH(255,239), and CCSDS LDPC(128,64) \cite{channelcodes} are discussed below, and we mainly focus on FER or BER performance of parallelizable decoders. The choice of $I_m$ assumes negligible FER improvement is expected when further increasing it for each code and decoder, thus specified case-by-case. For the proposed QBP or QBP-SF prepared for all BCH codes, $(\delta_1, \delta_2) = (9,2)$ are fixed throughout for convenience. Source code of their implementation is available on GitHub\footnote{\url{https://github.com/lgw-frank/Short\_BCH\_quasi-BP\_decoding}} for easy reproduction of the simulations.

\subsection{Decoding Performance}

Since LDPC(128,64) and BCH(127,64) codes share similar blocklength and rate, the former serves as a suitable reference for the latter regarding FER comparison of various decoders.

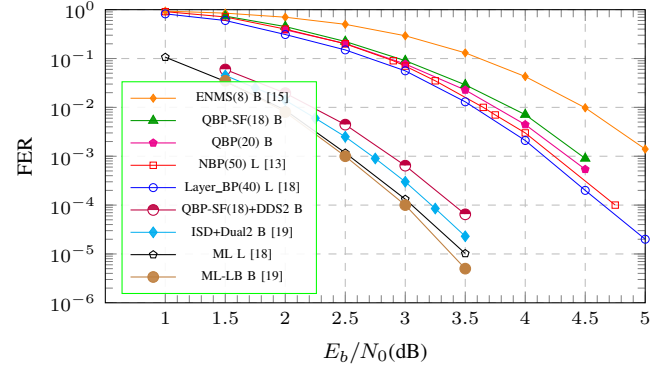
\begin{figure}[htbp]
	\centering
	\input{plots/BCH_LDPC_FER}   
	\caption{Decoding performance of BCH(127,64) ('B') and LDPC(128,64) ('L') codes with multiple decoders.}
	\label{fig_bch_ldpc_fer_ber}
\end{figure}

As shown in Fig.~\ref{fig_bch_ldpc_fer_ber}, for the BCH(127,64) code, the proposed QBP-SF surpasses ENMS by more than 0.6 dB at the cost of both higher $I_m$ (18 versus 8) and computational complexity, while the former lags behind QBP with $I_m=20$ by about 0.1 dB. For the LDPC (128,64) code, layered BP with $I_m=40$ provides 0.1 dB gain over the NBP decoder \cite{buchberger21} with $I_m=50$, due to the advantage of layered message scheduling over the flooding schedule of the latter. Notably, the decoding gap between QBP for the BCH code and layered BP for the LDPC code is within 0.25 dB. It is also observed that these decoders, whether for BCH or LDPC codes, exhibit nearly identical slopes, indicating that the FER gap between each other will not widen within the rendered SNR region. However, in the extreme high-SNR region where LDPC codes are known to suffer from error floor \cite{richardson2003error}, BCH codes are a better option since their dense structure naturally prevents the formation of trapping sets that are the source of the error floor.

Furthermore, the hybrid of QBP-SF and DDS2 (an OSD variant with a fixed list of $10^4$ TEPs \cite{li2025neural}) catches up most of the significant gap between iterative decoders and ML, actually within 0.4 dB. Although the hybrid lags about 0.15 dB behind the 'ISD+Dual2' \cite{bossert2022hard}, which is an OSD variant with $10^4$ TEPs, the latter suffers from customized processing of each received sequence, hence limited throughput.

\begin{figure}[htbp]
	\centering
	\resizebox{\linewidth}{!}{\input{plots/BCH_127_99_255_239}}
	\caption{FER/BER (solid/dashed lines) for BCH(127,99) ('B1') and (255,239) ('B2') codes with multiple decoders.}
	\label{fig_127_99_255_239_bch_fer}
\end{figure}
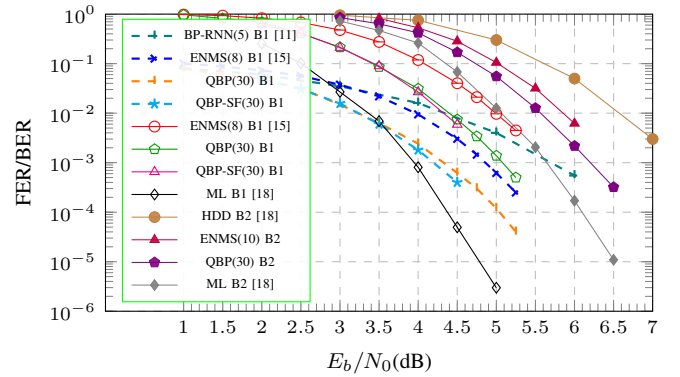

For the higher-rate BCH (127,99) and (255,239) codes, the FER and BER performance of various decoders are shown in Fig.~\ref{fig_127_99_255_239_bch_fer}.

In terms of BER (dashed lines) for the BCH(127,99) code, BP-RNN \cite{nachmani18}, as one variant of NBP, lags behind ENMS by at least 0.6 dB at $\mathrm{BER} = 10^{-3}$, and the gap continues to widen with increased SNR due to differing curve slopes. Both QBP and QBP-SF with $I_m=30$ show another 0.5 dB improvement over ENMS at the cost of complexity. The performance discrepancy in terms of FER corroborates the observations on the BER metric. For the BCH(255,239) code, QBP (with QBP-SF omitted for clarity) is only about 0.6 dB away from ML performance at FER = $10^{-3}$, and due to similar slopes, this gap is not expected to increase with SNR. In comparison, ENMS ends up about 0.3 dB behind QBP; hard-decision decoding (HDD), operating in serial mode, lags behind QBP by about 1.0 dB, despite its lightest complexity. Furthermore, for both high-rate BCH codes, the concatenation of QBP (or QBP-SF) with an OSD variant enables achieving the ML bound more readily, considering the performance gap between them is below 1.0 dB; this is omitted in Fig.~\ref{fig_127_99_255_239_bch_fer} to avoid cluttering.

\subsection{Complexity Analysis of Check Node Update}

Let $F(\cdot)$ denote the number of floating-point operations required for a given operation. All arithmetic operations, besides comparison, count as one FLOP each. The $\phi$ function in the check update of QBP is defined as:
$\phi(x) = -\log\left(\tanh\left(\frac{x}{2}\right)\right)$.
It is self-inverse and is used to transform messages to the $\phi$-domain, where linear operations can be performed, and $F(\phi)=40$ is assumed hereafter.

Ignoring sign processing, for the BCH(127,99) code with $r_w = 44$, $(2r_wF(\phi)+2r_w-1) = 3599$ FLOPs per check node are required in QBP. In comparison, for QBP-SF, given the setups in the NN architecture presented in Fig.~\ref{fig:nn_substitution}, where HardSwish counts as 5 FLOPs, the total FLOPs per check node is 9036 and the total number of trainable parameters is 997, as detailed in Table~\ref{NN_complexity}.

\begin{table}[h]
\centering
\caption{Layer-wise Complexity Breakdown of NN Component of QBP-SF for BCH(127,99) Code with $r_w=44$}
\label{NN_complexity}
\resizebox{0.48\textwidth}{!}{%
\begin{tabular}{lcccc}
\toprule
\textbf{Layer} & \textbf{Input Shape} & \textbf{Output Shape} & \textbf{Params} & \textbf{FLOPs} \\
\midrule
Conv1 & (44, 1) & (42, 8) & 32 & 3,024 \\
Conv2 & (42, 8) & (40, 4) & 100 & 4,800 \\
Flatten & (40, 4) & (160) & 0 & 0 \\
Dense1 & (160) & (4) & 644 & 664 \\
Dense2 & (4) & (44) & 220 & 220 \\
Scale & (44) & (44) & 1 & 44 \\
ReLU & (44) & (44) & 0 & 44 \\
Sort & (44) & (44) & 0 & 240 \\
\midrule
\textbf{Total} & - & - & \textbf{997} & \textbf{9,036} \\
\bottomrule
\end{tabular}%
}
\end{table}

The NN's reliance on multiply-accumulate operations aligns well with optimized hardware accelerators. In contrast, the $\phi$ function in QBP requires specialized hardware support or extensive look-up tables, making it less hardware-friendly despite having fewer FLOPs. By comparison, ENMS requires substantially less complexity, i.e., $4r_w = 176$ FLOPs per check node, but at the cost of clear decoding degradation.

%% file: plots/BCH_LDPC_FER.tex
\begin{tikzpicture}
	\begin{semilogyaxis}[
		width=1.4\textwidth, 
		height=0.8\textwidth,      
		scale = 0.3,
		label style={font=\footnotesize},      
		tick label style={font=\scriptsize},      
		xlabel={$E_b/N_0$(dB)},
		ylabel={FER},
		xmin=0.5, xmax=5.0,
		ymin=1e-6, ymax=1,
		xtick={1.0,1.5,2,2.5,3,3.5,4,4.5,5.0},
		ytick={1e-6,1e-5,1e-4,1e-3,1e-2,1e-1,1},
		legend pos = south west,
		legend columns=1,
        ymajorgrids=true,
		xmajorgrids=true,
		minor x tick num=4,
		minor grid style={dotted, gray!50},
		grid style=dashed,
		legend style={font=\tiny\selectfont, fill opacity=0.9,
        text opacity=1},
		]
		\addplot[
		color=orange,
		mark=diamond*,
		mark size=1.5pt,
		very thin
		] coordinates {
			(1.00,0.93495)
			(1.50,0.84881)
			(2.00,0.70094)
			(2.50,0.50150)
			(3.00,0.29223)
			(3.50,0.13072)
			(4.00,0.0428)
			(4.50,0.0098)
			(5.00,0.0014)
		};
		\addlegendentry{ENMS(8) B \cite{li2025effective}}
		
		\addplot[
		color=green!60!black,
		mark=triangle*,
		solid,
		thin
		]
		coordinates {
			(1.50, 0.73738)
			(2.00, 0.45081)
			(2.50, 0.22316)
			(3.0, 0.09034)
			(3.50,0.02894)
			(4.00,0.00710)
			(4.5,0.0009)
		};	
		\addlegendentry{QBP-SF(18) B}
		
        \addplot[
		color=magenta,
		mark=pentagon*,
		mark size=1.5pt,
		thin
		] coordinates {
			(2.00,0.38692)
			(2.50,0.2024)
			(3.00,0.07708)
			(3.50,0.02269)
			(4.00,0.00444)
			(4.50,0.00054)
		};
		\addlegendentry{QBP(20) B}
		
		\addplot[
		color=red,
		mark=square,
		mark size=1.2pt,
		very thin
		] coordinates {
			(1.0,0.91)
			(1.5,0.71)
			(2.0,0.4)
			(2.5,0.2)
			(2.9,0.09)
			(3.0,0.07)
			(3.25,0.035)
			(3.65,0.01)
			(3.75,0.007)
			(4.0,0.003)
			(4.75,1e-4)
		};
		\addlegendentry{NBP(50) L \cite{buchberger21}}
		
		\addplot[
		color=blue,
		mark=o,
		mark size=1.5pt,
		very thin
		] coordinates {
			(1.0,0.82)
			(1.5,0.6)
			(2.0,0.31)
			(2.5,0.15)
			(3.0,0.056)
			(3.5,0.013)
			(4.0,0.0021)
			(4.5,2e-4)
			(5.0,2e-5)
		};
		\addlegendentry{Layer\_BP(40) L \cite{channelcodes}}
		
		\addplot[
		color=purple,
		mark=halfcircle*,
		solid,
		thin
		]
		coordinates {
			(1.5, 0.0595803)
			(2.0, 0.0195652)
			(2.5, 0.0044327)
			(3.0, 0.0006414)
			(3.5, 0.0000647)
		};	
		\addlegendentry{QBP-SF(18)+DDS2 B}
		
		\addplot[
		color=cyan,
		mark=diamond*,
		solid,
		very thin
		]
		coordinates {
			(1.5, 0.045)
			(1.75,0.025)
			(2.0, 0.014)
			(2.25,6e-3)
			(2.5, 2.5e-3)
			(2.75,9e-4)
			(3.0, 3e-4)
			(3.25,8.5e-5)
			(3.5, 2.3e-5)
		};	
		\addlegendentry{ISD+Dual2 B\cite{bossert2022hard}}
		
        \addplot[
		color=black,
		mark=pentagon,
		mark size=1.5pt,
		thin
		] coordinates {
			(1.00, 1.064e-1)
			(1.50, 3.397e-2)
			(2.00, 8.773e-3)
			(2.50, 1.168e-3)
			(3.00, 1.321e-4)
			(3.50, 1.022e-5)
		};	
		\addlegendentry{ML L \cite{channelcodes}}
		
		\addplot[
		color=brown,
		mark=*,
		solid,
		thin
		]
		coordinates {
			(1.5,3.5e-2)
			(2.00, 8e-03)
			(2.50, 1.0e-03)
			(3.00, 1.0e-04)
			(3.50, 5.0e-06)
		};	
		\addlegendentry{ML-LB B\cite{bossert2022hard}}

    \end{semilogyaxis}
\end{tikzpicture}

%% file: plots/BCH_127_99_255_239.tex
\centering
\begin{tikzpicture}
\begin{semilogyaxis}[
		width=1.4\textwidth, 
		height=0.8\textwidth,      
		scale = 0.3,
		label style={font=\footnotesize},      
		tick label style={font=\scriptsize},      
		xlabel={$E_b/N_0$(dB)},
		ylabel={FER/BER},
		xmin=0.0, xmax=7.0,
		ymin=1e-6, ymax=1,
		xtick={1.0,1.5,2,2.5,3,3.5,4,4.5,5.0,5.5,6.0,6.5,7.0},
		ytick={1e-6,1e-5,1e-4,1e-3,1e-2,1e-1,1},
		legend pos = south west,
		legend columns=1,
        ymajorgrids=true,
		xmajorgrids=true,
		minor x tick num=4,
		minor grid style={dotted, gray!50},
		grid style=dashed,
		legend style={font=\tiny\selectfont, fill opacity=0.9,
        text opacity=1},
		]
		\addplot[
		color=teal,
		mark=+,
        thick,
		dashed
		]
		coordinates {
			(1.0,0.08)
			(2.0,0.06)
			(3.0,0.035)
			(4.0,0.016)
			(5.0,0.004)
			(6.0,5.5e-4)
		};
		\addlegendentry{BP-RNN(5) B1 \cite{nachmani18}}
		
		\addplot[
		color=blue,
		mark=x,
        thick,
		dashed,
		]
		coordinates {
			(1.00,0.10024)(1.50,0.08735)(2.00,0.07330) (2.50,0.05643) (3.00,0.03756) (3.50,0.02182) (4.00,0.00945) (4.50,0.00303)
			(4.75, 0.00145)
			(5.00, 0.00062)
			(5.25, 0.00025)
		};	
		\addlegendentry{ENMS(8) B1 \cite{li2025effective}}
		
		\addplot[
		color=orange,
		mark=|,
        thick,
		dashed]
		coordinates {
			(1.00,0.07816)
			(1.50,0.05963)
			(2.00,0.04903)
			(2.50,0.03146)
			(3.00,0.01490)
			(3.50,0.00632)
			(4.00,0.00241)
			(4.50,0.00062)
			(4.75, 0.0003)
			(5.00,0.00012)
			(5.25,0.00004)
		};	
		\addlegendentry{QBP(30) B1}
		
	\addplot[
		color=cyan,
		mark=star,
        thick,
		dashed
	]
	coordinates {
		(2.00,0.0484)
		(2.50,0.0305)
		(3.00,0.0157)
		(3.50,0.0060)
		(4.00,0.0018)
		(4.50,0.0004)
	};	
	\addlegendentry{QBP-SF(30) B1}
	
		\addplot[
		color=red,
		mark=halfcircle,
		thin
		]
		coordinates {
			(1.00, 0.97777)
			(1.50, 0.93065)
			(2.00, 0.82760)
			(2.50, 0.68020)
			(3.00, 0.47643)
			(3.50, 0.27441)
			(4.00, 0.11829)
			(4.50, 0.04015)
			(4.75, 0.02102)
			(5.00, 0.00969)
			(5.25, 0.00447)
		};	
		\addlegendentry{ENMS(8) B1 \cite{li2025effective}}
		
		\addplot[
		color=green!60!black,
		mark=pentagon,
		thin
		]
		coordinates {
			(1.00,0.96667)
			(1.50,0.82000)
			(2.00,0.68667)
			(2.50,0.40800)
			(3.00,0.21400)
			(3.50,0.08583)
			(4.00,0.03118)
			(4.50,0.00745)
			(4.75,0.0034)
			(5.00, 0.00138)
			(5.25,0.0005)
		};	
		\addlegendentry{QBP(30) B1}

		\addplot[
		color=magenta,
		mark=triangle,
		thin
		]
		coordinates {
			(2.00,0.6056)
			(2.50,0.4111)
			(3.00,0.2104)
			(3.50,0.0895)
			(4.00,0.0267)
			(4.50,0.0059)
		};	
		\addlegendentry{QBP-SF(30) B1}
		
		\addplot[
		color=black,
		mark=diamond,
        thin
		]
		coordinates {
			(2.00, 2.551e-01)
			(2.50, 1.020e-01)
			(3.00, 2.680e-02)
			(3.50, 6.907e-03)
			(4.00, 8.074e-04)
			(4.50, 4.958e-05)
			(5.00, 2.984e-06)
		};	
		\addlegendentry{ML B1 \cite{channelcodes}}
		
		\addplot[
		color=brown,
		mark=*,
		very thin
		]
		coordinates {
			(3.0,0.95)
			(4.0,0.75)
			(5.0,0.3)
			(6.0,0.05)
			(7.0,3e-3)
			(8.0,6e-5)
			(8.5,6e-6)
		};
		\addlegendentry{HDD B2 \cite{channelcodes}}
\addplot[
color=purple,
mark=triangle*,
very thin
]
coordinates {
(3.50,0.8175)
(4.00,0.5300)
(4.50,0.2836)
(5.00,0.1052)
(5.50,0.0317)
(6.00,0.0062)
};	
\addlegendentry{ENMS(10) B2}		
		\addplot[
		color=violet,
		mark=pentagon*,
		very thin
		]
		coordinates {
			(3.00,0.85000)
			(3.50,0.63636)
			(4.00,0.41875)
			(4.50,0.16917)
			(5.00,0.05537)
			(5.50,0.01267)
			(6.00,0.00218)
			(6.50,3.2e-4)
		};	
		\addlegendentry{QBP(30) B2}

		\addplot[
		color=gray,
		mark=diamond*,
        very thin
		]
		coordinates {
			(3.00, 7.246e-01)
			(3.50, 4.673e-01)
			(4.00, 2.604e-01)
			(4.50, 6.793e-02)
			(5.00, 1.256e-02)
			(5.50, 2.054e-03)
			(6.00, 1.708e-04)
			(6.50, 1.093e-05)
		};	
		\addlegendentry{ML B2 \cite{channelcodes}}
    \end{semilogyaxis}
\end{tikzpicture}

%% file: sections/conclusions.tex
\section{Conclusions}
\label{conclusions}
This paper demonstrates that BP-like parallel decoding schemes---namely QBP and QBP-SF---can be designed for BCH codes to achieve FER performance competitive with traditional BP variants, which have conventionally been considered effective only for LDPC codes. In the QBP decoder, the variable and check node update rules of standard BP are largely retained, with key modifications including the dilation of inputs via BCH code automorphisms, the optimization and deliberate augmentation of the parity-check matrix with redundant rows, and the normalization of a posteriori messages after each iteration. The QBP-SF decoder further replaces the nonlinear $\tanh$ and $\tanh^{-1}$ functions in the check node update with a neural network, while preserving the remaining structure of QBP.

A fundamental distinction between QBP-SF and conventional BP variants lies in the processing of incoming messages at the check nodes: the fixed composition of $\tanh$ and $\tanh^{-1}$ is substituted by a flexible mapping $f_{\boldsymbol{\theta}}$ parameterized by learnable weights $\boldsymbol{\theta}$. This neural parameterization endows QBP-SF with the potential to learn optimal message update rules, particularly in non-AWGN channel environments, an avenue that merits further exploration.

Notably, the proposed check-node substitution using neural networks can be naturally extended to BP decoding of LDPC codes. Given the typically much lower row weights of LDPC codes, the required neural network can be designed with fewer layers or filters, rendering the approach a compelling alternative in scenarios where standard BP and simplified versions such as NMS decoders exhibit a clear FER performance gap, e.g., in the decoding of low-rate LDPC codes.

\section*{Acknowledgments}
Part of the material presented in this paper is the subject of a pending patent application in China.

%% file: main.bbl
\begin{thebibliography}{10}
\providecommand{\url}[1]{#1}
\csname url@samestyle\endcsname
\providecommand{\newblock}{\relax}
\providecommand{\bibinfo}[2]{#2}
\providecommand{\BIBentrySTDinterwordspacing}{\spaceskip=0pt\relax}
\providecommand{\BIBentryALTinterwordstretchfactor}{4}
\providecommand{\BIBentryALTinterwordspacing}{\spaceskip=\fontdimen2\font plus
\BIBentryALTinterwordstretchfactor\fontdimen3\font minus \fontdimen4\font\relax}
\providecommand{\BIBforeignlanguage}[2]{{%
\expandafter\ifx\csname l@#1\endcsname\relax
\typeout{** WARNING: IEEEtran.bst: No hyphenation pattern has been}%
\typeout{** loaded for the language `#1'. Using the pattern for}%
\typeout{** the default language instead.}%
\else
\language=\csname l@#1\endcsname
\fi
#2}}
\providecommand{\BIBdecl}{\relax}
\BIBdecl

\bibitem{shannon1948mathematical}
C.~E. Shannon, ``A mathematical theory of communication,'' \emph{Bell Syst. Tech. J.}, vol.~27, no.~3, pp. 379--423, 1948.

\bibitem{gallager62}
R.~Gallager, ``Low-density parity-check codes,'' \emph{IRE Trans. Inf. Theory}, vol.~8, no.~1, pp. 21--28, 1962.

\bibitem{mackay96}
D.~J. MacKay and R.~M. Neal, ``Near shannon limit performance of low density parity check codes,'' \emph{Electron. Lett.}, vol.~32, no.~18, p. 1645, 1996.

\bibitem{zhao05}
J.~Zhao, F.~Zarkeshvari, and A.~H. Banihashemi, ``On implementation of {Min-Sum} algorithm and its modifications for decoding low-density parity-check ({LDPC}) codes,'' \emph{IEEE Trans. Commun.}, vol.~53, no.~4, pp. 549--554, 2005.

\bibitem{chen2005reduced}
J.~Chen, A.~Dholakia, E.~Eleftheriou, M.~P. Fossorier, and X.-Y. Hu, ``Reduced-complexity decoding of {LDPC} codes,'' \emph{IEEE Trans. Commun.}, vol.~53, no.~8, pp. 1288--1299, 2005.

\bibitem{richardson2003error}
T.~Richardson, ``Error floors of {LDPC} codes,'' in \emph{Proc. Annu. Allerton Conf. Commun. Control Comput.}, vol.~41, 2003, pp. 1426--1435.

\bibitem{zhang2024improved}
Z.~Zhang, K.~Niu, and H.~Cui, ``Improved {BCH}-{P}olar concatenated scheme for unsourced random access in {Internet of Things},'' \emph{{IEEE} Internet Things J.}, vol.~11, no.~19, pp. 32\,172--32\,182, 2024.

\bibitem{Fossorier1995}
M.~P. Fossorier and S.~Lin, ``Soft-decision decoding of linear block codes based on ordered statistics,'' \emph{IEEE Trans. Inf. Theory}, vol.~41, no.~5, pp. 1379--1396, 1995.

\bibitem{Yue2021}
C.~Yue, M.~Shirvanimoghaddam, G.~Park, O.-S. Park, B.~Vucetic, and Y.~Li, ``Probability-based ordered-statistics decoding for short block codes,'' \emph{IEEE Commun. Lett.}, vol.~25, no.~6, pp. 1791--1795, 2021.

\bibitem{nachmani16}
E.~Nachmani, Y.~Be'ery, and D.~Burshtein, ``Learning to decode linear codes using deep learning,'' in \emph{Allerton Conf. Commun., Control, Comput.}\hskip 1em plus 0.5em minus 0.4em\relax IEEE, 2016, Conference Proceedings, pp. 341--346.

\bibitem{nachmani18}
E.~Nachmani, E.~Marciano, L.~Lugosch, W.~J. Gross, D.~Burshtein, and Y.~Be'ery, ``Deep learning methods for improved decoding of linear codes,'' \emph{IEEE J. Sel. Top. Signal Process.}, vol.~12, no.~1, pp. 119--131, 2018.

\bibitem{liang18}
F.~Liang, C.~Shen, and F.~Wu, ``An iterative {BP-CNN} architecture for channel decoding,'' \emph{IEEE J. Sel. Top. Signal Process.}, vol.~12, no.~1, pp. 144--159, 2018.

\bibitem{buchberger21}
A.~Buchberger, C.~Häger, H.~D. Pfister, L.~Schmalen, and A.~G. i~Amat, ``Learned decimation for neural belief propagation decoders,'' in \emph{IEEE Int. Conf. Acoust., Speech, Signal Process. (ICASSP)}.\hskip 1em plus 0.5em minus 0.4em\relax IEEE, 2021, Conference Proceedings, pp. 8273--8277.

\bibitem{von2024spiking}
A.~von Bank, E.-M. Edelmann, S.~Miao, J.~Mandelbaum, and L.~Schmalen, ``Spiking neural belief propagation decoder for short block length {LDPC} codes,'' \emph{{IEEE} Commun. Lett.}, vol.~29, no.~1, pp. 45--49, 2024.

\bibitem{li2025effective}
G.~Li and X.~Yu, ``Effective application of normalized {Min-Sum} decoding for short {BCH} codes,'' \emph{{IEEE} Commun. Lett.}, vol.~29, no.~8, pp. 1983--1987, 2025.

\bibitem{li2025neural}
------, ``Neural-model-augmented hybrid {NMS}-{OSD} decoders for near-{ML} in short block codes,'' \emph{arXiv preprint arXiv:2509.25580}, 2025.

\bibitem{kabat2007new}
A.~Kabat, F.~Guilloud, and R.~Pyndiah, ``New approach to order statistics decoding of long linear block codes,'' in \emph{IEEE GLOBECOM}, 2007, pp. 1467--1471.

\bibitem{channelcodes}
M.~Helmling, S.~Scholl, F.~Gensheimer, T.~Dietz, K.~Kraft, O.~Griebel, S.~Ruzika, and N.~Wehn, ``{D}atabase of {C}hannel {C}odes and {ML} {S}imulation {R}esults,'' \url{www.rptu.de/channel-codes}, 2025.

\bibitem{bossert2022hard}
M.~Bossert, R.~Schulz, and S.~Bitzer, ``On hard and soft decision decoding of {BCH} codes,'' \emph{{IEEE} Trans. Inf. Theory}, vol.~68, no.~11, pp. 7107--7124, 2022.

\end{thebibliography}
